\documentclass[a4paper]{JHEP3}

\usepackage{graphicx}
\usepackage{amsmath}
\usepackage{cite}

\def\eqref#1{(\ref{#1})}
\def\text{\rm }
\def\Res{{\rm Res}}
\def\samurai{{{\sc samurai}}}

\newcommand{\beq}{\begin{equation}}
\newcommand{\eeq}{\end{equation}}
\newcommand{\bqa}{\begin{eqnarray}}
\newcommand{\eqa}{\end{eqnarray}}

\def\db#1{\bar D_{#1}}

\def\slh#1{\rlap / {#1}}
\def\eqn#1{Eq.~(\ref{#1})}

\def\fig#1{Fig.~{\ref{#1}}}

\def\spa#1.#2{\langle#1\,#2\rangle}
\def\spb#1.#2{[#1\,#2]}

\def\spab#1.#2.#3{\langle\mskip-1mu{#1}
                  | #2 | {#3}]}

\def\spba#1.#2.#3{[\mskip-1mu{#1}
                  | #2 | {#3}\rangle}

\def\spbb#1.#2.#3.#4{[\mskip-1mu{#1}
                     | {#2} \ {#3} | {#4}]}

\def\spaa#1.#2.#3.#4{\langle\mskip-1mu{#1}
                     | {#2} \ {#3} | {#4}\rangle}

\newcommand{\bea}{\begin{eqnarray}}

\newcommand{\eea}{\end{eqnarray}}

\newcommand{\bean}{\begin{eqnarray*}}

\newcommand{\eean}{\end{eqnarray*}}
           
\newcommand{\nn}{\nonumber \\}

\title{Scattering AMplitudes from Unitarity-based \\ Reduction Algorithm 
at the Integrand-level}

\author{P. Mastrolia \\
Centro Studi e Ricerche ``E. Fermi'', Piazza del Viminale 1,
I-00184, Rome, Italy.\\
Dipartimento di Fisica, Universit\`a di Salerno, via Ponte don Melillo,
I-84084, Fisciano (SA), Italy.\\
Theory Group, Physics Department, CERN,
           CH-1211 Geneva 23, Switzerland.\\ 
E-mail: \email{pierpaolo.mastrolia@cern.ch}
}    

\author{G. Ossola \\ 
Physics Department, New York City College of Technology,\\ 
                      City University Of New York, 
           300 Jay Street, Brooklyn NY 11201, USA.\\
E-mail: \email{GOssola@citytech.cuny.edu}
}

\author{T. Reiter \\
Nikhef, Science Park 105, 1098XG Amsterdam, The Netherlands. \\
E-mail: \email{thomasr@nikhef.nl}
}

\author{F. Tramontano \\
Theory Group, Physics Department, CERN,
           CH-1211 Geneva 23, Switzerland.\\
E-mail: \email{francesco.tramontano@cern.ch}
}

\preprint{CERN-PH-TH/2010-124 \\ Nikhef-2010-015}

\abstract{
{\samurai} is a tool for the automated numerical evaluation 
of one-loop corrections
to any scattering amplitudes within the dimensional-regularization scheme. 
It is based on the decomposition of the integrand according to the 
OPP-approach, 
extended to accommodate an implementation of the generalized $d$-dimensional 
unitarity-cuts technique, 
and uses a polynomial interpolation exploiting the 
Discrete Fourier Transform.
{\samurai} can process integrands written either as numerator 
of Feynman diagrams or as product of tree-level amplitudes. 
We discuss some applications, among which the 6- and 8-photon 
scattering in QED, and the 6-quark scattering in QCD.
{\samurai} has been
implemented as a {\tt Fortran90} library, publicly available,   
and it could be a useful module 
for the systematic evaluation of the virtual corrections
oriented towards automating next-to-leading order calculations relevant for 
the LHC phenomenology.
}

\begin{document}

\section{Introduction}

With the beginning of the experimental programs at the LHC, 
the need of describing particle scattering events with high accuracy 
becomes more pressing. On the theoretical side, perturbative calculation within leading order precision 
cannot be sufficient, therefore accounting for effects 
due to next-to-leading order corrections becomes mandatory.

Leading order (LO) processes are the core of well-established multi-purpose event generators
like MadGraph MadEvent
\cite{Stelzer:1994ta,Maltoni:2002qb,Alwall:2007st}, 
CompHEP-CalcHEP
\cite{Boos:2004kh,Pukhov:2004ca}, 
SHERPA
\cite{Gleisberg:2003xi,Gleisberg:2008ta}, 
WHIZARD
\cite{Kilian:2007gr},
ALPGEN \cite{Mangano:2002ea}, and HELAC \cite{Kanaki:2000ey,Cafarella:2007pc},
whereas a variety of processes computed at NLO are currently implemented in programs 
like MCFM \cite{Campbell:1999ah,Campbell:2002tg} and NLOJET++ \cite{Nagy:2003tz},
or MC@NLO \cite{Frixione:2002ik,Frixione:2006gn} and the POWHEG
\cite{Nason:2004rx,Nason:2006hfa,LatundeDada:2006gx,Frixione:2007nw,Alioli:2008gx,Hamilton:2008pd,Alioli:2010xd} 
which consider also the matching with parton showers.

The next-to-leading order (NLO) corrections to an $n$-parton 
final state process 
receive contributions from two sources: the one-loop
correction to the $(2\to n)$-scattering, due to the exchange 
of an internal virtual particle; and the 
tree-level scattering $(2\to n+1)$, due to the real emission of 
an extra parton. 
Each contribution contains divergencies which  
cancel mutually in the final result where they are combined.

The extraction of the real radiation singularities from general
processes has been addressed with techniques based either on phase-space
slicing~\cite{Giele:1991vf} or on the use of universal
subtraction terms~\cite{Ellis:1980wv,Kunszt:1992tn}, which at present have been 
implemented in several algorithms, like the 
FKS subtraction~\cite{Frixione:1995ms}, 
dipole subtraction~\cite{Catani:1996vz,Catani:2002hc,Gleisberg:2007md,Seymour:2008mu,Czakon:2009ss,Hasegawa:2009tx,Frederix:2008hu,Frederix:2009yq,Frederix:2010cj} 
and antenna subtraction~\cite{Kosower:1997zr,Campbell:1998nn,GehrmannDeRidder:2005cm,Daleo:2006xa}.

The increasing computational complexity of one-loop amplitudes, 
when the number of particles involved in the scattering increases, 
has limited the possibility of developing an automated multi-process evaluator
for scattering amplitudes at NLO. The available results have been so far computed on a 
process-by-process basis, but, due to the recent advances in computational 
techniques for high-energy physics, that possibility is now at the horizon. 

Currently, the state-of-the-art is represented by the numerical 
calculation of extremely challenging $2\to 4$ processes, like 
the EW corrections to $e^+ e^- \to 4 {\rm f}$~\cite{Denner:2005nn,Denner:2005fg},
or the NLO QCD corrections to  $pp\to W+3$~jet
production~\cite{Berger:2009zg,Berger:2009ep,Ellis:2009zw,KeithEllis:2009bu},
$pp\to Z+3$~jet production~\cite{Berger:2010vm},
$pp\to t\bar t b\bar b$~\cite{Bredenstein:2009aj,Bredenstein:2010rs,Bevilacqua:2009zn} and
$pp\to t\bar t jj$~\cite{Bevilacqua:2010ve},
and $q\bar q \to b\bar b b\bar b$~\cite{Binoth:2009rv},
which have been obtained both by optimizing the algebraic tensor reduction, 
and by developing novel approaches based 
on properties of scattering amplitudes such as factorization and unitarity. 
Also, the development of novel analytic techniques has benefited 
from a more systematic use of unitarity-based methods in combination with
the theory of multivariate-complex functions \cite{Britto:2004nc,Britto:2005ha,Britto:2006sj,Mastrolia:2006ki,Forde:2007mi,BjerrumBohr:2007vu,Kilgore:2007qr,Badger:2008cm,Mastrolia:2009dr,Mastrolia:2009rk}, yielding the recent completion 
of the one-loop QCD correction to $pp \to H+2$ jets in the heavy-top 
limit \cite{Berger:2006sh,Badger:2006us,Badger:2007si,Glover:2008ffa,Badger:2009hw,Dixon:2009uk,Badger:2009vh}.

It is well known that any one-loop amplitude can be 
expressed as a linear combination of a limited set of 
Master Integrals (MI) \cite{Passarino:1978jh,'tHooft:1978xw}:
therefore, the evaluation of one-loop corrections reduces to evaluating 
the coefficients that multiply each MI.
Aiming at the full reconstruction of one-loop amplitudes 
through such a decomposition, several automated packages have appeared, 
either in public releases like
CutTools~\cite{Ossola:2007ax} and Golem~\cite{Binoth:2008uq},
or in private versions such as the routines described in \cite{Denner:2005nn}, 
\cite{Lazopoulos:2008ex} and \cite{Winter:2009kd},
and codes like BlackHat~\cite{Berger:2008sj}, Rocket~\cite{Giele:2008bc}, 
and Helac-1Loop~\cite{vanHameren:2009dr}.

The development of novel numerical techniques have received a boost 
by the combination of three important ideas:
\begin{itemize}
\item[{\it i})]
 universal four-dimensional decomposition for the numerator of 
the integrand for any one-loop scattering amplitudes 
\cite{delAguila:2004nf,Ossola:2006us}; 
\item[{\it ii})]
four-dimensional unitarity-cuts, detecting only 
the (poly)logarithmic structure of the 
amplitude, known as the cut-constructible term \cite{Bern:1994zx,Britto:2004nc} 
(see \cite{Berger:2009zb} for a more comprehensive list of references); 
\item[{\it iii})]
 unitarity-cuts in $d$-dimension, yielding the complete
determination of dimensionally regulated one-loop amplitudes \cite{Mahlon:1993fe,Bern:1995db,Anastasiou:2006jv,Anastasiou:2006gt,Giele:2008ve,Ellis:2008ir}.
\end{itemize}
The first two ideas merged in the by-now known as 
OPP-approach \cite{Ossola:2006us,Ossola:2007bb}, 
proposed by Papadopoulos, Pittau, and one of us, 
where the multi-pole decomposition 
of the numerator of any Feynman integral is achieved by a polynomial sampling 
that exploits the solutions of generalized unitarity-cuts.

In the context of four-dimensional unitarity, 
the problem of computing the cut constructible
term and the rational term, that escapes the four-dimensional detection, 
are necessarily considered as separate issues. The reconstruction of 
the latter usually requires information from an extra source.
When not obtained from the direct calculation of Feynman integrals, 
the rational term 
can be reconstructed by adding a piece derived  
from the cut-constructible 
part (for instance, the overlapping-term within the on-shell 
method \cite{Bern:2005cq}, 
or the $R_1$-term within the OPP-approach \cite{Ossola:2008xq}),
and a remaining piece computed through an additional tree-level like  
construction (for instance, the BCFW-recursive term within the on-shell method \cite{Bern:2005cq}, 
or the $R_2$-term within the OPP-approach \cite{Ossola:2008xq,Draggiotis:2009yb,Garzelli:2009is}).

The idea of performing unitarity-cuts in $d$-dimension \cite{Mahlon:1993fe,Bern:1995db,Anastasiou:2006jv,Anastasiou:2006gt,Giele:2008ve,Ellis:2008ir,Badger:2008cm} yields a combined determination of both 
cut-constructible and rational terms at once. This technique has been neatly systematized 
for numerical purposes by Ellis, Giele, Kunszt and Melnikov \cite{Giele:2008ve,Ellis:2008ir}, 
and later proposed also for the on-shell approach by Badger \cite{Badger:2008cm}. 

\bigskip 

In this paper we present {\samurai}, a tool based on a hybrid algorithm 
for the numerical computation of one-loop amplitudes.
{\samurai} relies on the extension of the OPP-polynomial structures
to include an explicit dependence on the extra-dimensional parameter 
needed for the automated computation of the full rational term 
according to the $d$-dimensional approach,
and  makes use of a polynomial interpolation based on the type of 
Discrete Fourier Transform (DFT) described in \cite{Mastrolia:2008jb}.

We aim at producing a versatile code which could deal with 
any one-loop corrections, in massless as well as massive theories.
Our reduction algorithm can process both (numerator of) Feynman
integrals, proper of diagrammatic methods,
and products of tree-level amplitudes, as adopted in the framework of unitarity-based techniques. \\
For a complete reconstruction of the rational term,
the input should contain an explicit dependence on 
the dimensional-regularization parameters. 
In fact, 
it is expected to have a polynomial behavior in 
$\mu^2$, being $\mu$ the radial integration 
variable in the extra-dimensional subspace,
and in $\epsilon \ ( \ =\!(4-d)/2 \ )$ 
according to the choice of the regularization scheme. 
The result is given as Laurent expansion in $\epsilon$ up to the finite-order, 
and accounts for the full rational terms.

{\samurai} is implemented as a {\tt Fortran90} library, publicly available at 
the webpage:
\begin{center}
{\tt http://cern.ch/samurai}
\end{center}
and it is linked to OneLOop \cite{vanHameren:2009dr} and QCDLoop \cite{Ellis:2007qk} 
for the numerical evaluation of the MI. 
We applied it to a series of known processes, 
like the four-, six-photon and eight-photon scattering in QED, 
the QCD virtual corrections to 
Drell-Yan, to the leading-color amplitude for $V+1$jet production, 
to the six-quark scattering, 
$q_1 {\bar q}_1 \to q_2 {\bar q}_2 \ q_3 {\bar q}_3$, 
and to the contributions of the massive-scalar loop-diagrams to 
the all-plus helicity five- and six-gluon scattering. \\
In particular, for the virtual corrections to 
$q_1 {\bar q}_1 \to q_2 {\bar q}_2 \ q_3 {\bar q}_3$ \cite{Binoth:2009rv},
we also considered the reduction of automatically generated  
integrands, by interfacing {\samurai} 
with an infrastructure derived
from {\tt golem-2.0}~\cite{Binoth:2010pb}, 
which provides numerators of Feynman integrals.\\
These examples are thought to be used both as a guide to understand the {\samurai} framework, and as templates
to generate the codes for other calculations.

In the context of collaborations among different groups aiming 
at automated NLO calculations relevant for LHC phenomenology~\cite{Binoth:2010ra}, 
and, therefore, providing complementary structures to be interfaced~\cite{Binoth:2010xt},
{\samurai} could constitute the module for the systematic evaluation of the virtual corrections.

\bigskip 

The paper is organized as follows.
The reduction algorithm is discussed in Section 2;
Section 3 describes the key-points of the {\samurai} library, 
while a series of applications are illustrated in Section 4.
In Section 5, we resume our conclusions.

\section{Reduction Algorithm}
\label{sec:Reduction}

The reduction method is based on the general 
decomposition for the {\it integrand} of a generic  
one-loop amplitude, originally proposed by Papadopoulos, Pittau and one of us
~\cite{Ossola:2006us,Ossola:2007bb}, and later extended by 
Ellis, Giele, Kunszt and Melnikov \cite{Giele:2008ve,Ellis:2008ir}. 
Within the dimensional regularization scheme, any one-loop $n$-point amplitude 
can be written as
\bea
&& {\cal A}_n = \int d^d {\bar q} \ A(\bar q, \epsilon) \ , \nn 
&& A(\bar q, \epsilon)= \frac{{\cal N}({\bar q}, \epsilon)}{\db{0}\db{1}\cdots \db{n-1}} \ , \nn
&& \db{i} = ({\bar q} + p_i)^2-m_i^2 = 
            (q + p_i)^2-m_i^2-\mu^2, \qquad (p_0 \ne 0)\,.
\label{def:An}
\eea
We use a bar to denote objects living in $d=~4-2\epsilon$  
dimensions, following the prescription
\bea \label{eq:qmu}
 \slh{{\bar q}} = \slh{q} + \slh{\mu} \ , \quad {\rm with} \qquad 
{\bar q}^2= q^2 - \mu^2 \ . 
\eea
Also, we use the notation $f({\bar q})$ as short-hand notation for $f(q,\mu^2)$.

\subsection{Integrands}

{\samurai} can reduce integrands of one-loop amplitudes 
which can be defined in two ways, either as {\it numerator functions}
(sitting on products of denominators), 
or as {\it products of tree-level amplitudes} (sewn along cut-lines).
The former definition accommodates a reduction based on a diagrammatic
method, while the latter is proper of a unitarity-based technology.

According to the chosen dimensional regularization scheme,
the most general numerator of one-loop amplitudes
$\mathcal{N}(\bar{q}, \epsilon)$ can be thought as composed of three terms,
\begin{equation}
\mathcal{N}(\bar{q}, \epsilon) =
            N_0(\bar{q})
+ \epsilon  N_1(\bar{q})
+\epsilon^2 N_2(\bar{q}).
\label{eq:genericN}
\end{equation}
The coefficients of this $\epsilon$-expansion,  $N_0$, $N_1$ and $N_2$, are functions of
$q^\nu$ and $\mu^2$, therefore in our discussion,
except when a distinction between them is necessarily required, 
we will simply talk about $N$, giving as understood that the
same logic would apply to each of the three contributions~$N_i$.

\subsubsection{Decomposition}
According to \cite{Ossola:2006us,Ossola:2007bb},
the numerator $N({\bar q})$ can be expressed in terms of denominators $\db{i}$, as follows
\bea
\label{eq:2}
N({\bar q}) &=&
\sum_{i < \!< m}^{n-1}
          \Delta_{ i j k \ell m}({\bar q})
\prod_{h \ne i, j, k, \ell, m}^{n-1} \db{h} 
+\sum_{i < \!< \ell}^{n-1}
          \Delta_{ i j k \ell }({\bar q})
\prod_{h \ne i, j, k, \ell}^{n-1} \db{h} 
+ \nn     &+&
\sum_{i < \!< k}^{n-1}
          \Delta_{i j k}({\bar q})
\prod_{h \ne i, j, k}^{n-1} \db{h} 
+\sum_{i < j }^{n-1}
          \Delta_{i j}({\bar q}) 
\prod_{h \ne i, j}^{n-1} \db{h} 
+\sum_{i}^{n-1}
          \Delta_{i}({\bar q}) 
\prod_{h \ne i}^{n-1} \db{h} \ , \qquad
\label{def:OPP:deco}
\eea
where $ i < \!< m $ stands for a lexicographic ordering $i < j < k < \ell < m$.
The functions $\Delta({\bar q}) = \Delta(q,\mu^2)$ are polynomials in the components 
of $q$ and in $\mu^2$. 
By using the decomposition (\ref{def:OPP:deco}) in Eq.(\ref{def:An}), 
the multi-pole nature of the integrand of any one-loop $n$-point 
amplitude becomes trivially exposed,
\bea
A(\bar q) &=&
\sum_{i < \!< m}^{n-1}
         { \Delta_{ i j k \ell m}({\bar q}) \over 
           \db{i} \db{j} \db{k} \db{\ell} \db{m} } 
+
\sum_{i < \!< \ell}^{n-1}
         { \Delta_{ i j k \ell }({\bar q}) \over 
           \db{i} \db{j} \db{k} \db{\ell} } 
+
\sum_{i < \!<  k }^{n-1}
         { \Delta_{i j k}({\bar q}) \over 
           \db{i} \db{j} \db{k} }
+ \nn &+& 
\sum_{i < j }^{n-1}
         { \Delta_{i j}({\bar q}) \over 
           \db{i} \db{j} } 
+
\sum_{i}^{n-1}
         { \Delta_{i}({\bar q}) \over
           \db{i} } \ ,
\eea
which, as we will see, is responsible of the decomposition of any dimensional 
regulated one-loop amplitude
in terms of Master Integrals (MI) associated to 4-, 3-, 2-, and 1-point functions, 
respectively called boxes, triangles, bubbles, and tadpoles.

\subsection{Polynomial Structures and Discrete Fourier Transform}

The calculation of a generic scattering amplitude amounts 
to the problem of extracting 
the coefficients of multivariate polynomials, generated at every step of the multiple-cut analysis.
To determine these coefficients we implement a semi-numerical algorithm whose
main features are:
\begin{itemize}
\item the extension of the OPP-polynomials \cite{Ossola:2006us,Ossola:2007bb} for quadruple-, triple- and double-cut 
      to the framework of $d$-dimensional unitarity \cite{Giele:2008ve,Ellis:2008ir};
\item the parametrization of the residue of the quintuple-cut affecting only the 
      polynomial dependence on the extra-dimension scale \cite{Melnikov:2010iu};
\item the numerical sampling of the multiple-cut solutions according to the type of Discrete Fourier
      Transform described in \cite{Mastrolia:2008jb}.
\end{itemize}

\subsubsection{Polynomials}
\label{sec:Polynomials}

In this section we review the interpolation of 
the polynomial ${\Delta}({\bar q})$, appearing in Eq.(\ref{eq:genericN}), 
implemented in {\samurai}.

For each cut, we decompose $q$, namely the 4-dimensional part
of ${\bar q}$, into a specific basis of four massless vectors $e_i$ \cite{delAguila:2004nf,Ossola:2006us,Forde:2007mi},
\bea
 q = - p_0 + x_1 e_1 + x_2 e_2 + x_3 e_3 + x_4 e_4 \ ,
\label{eq:q_deco}
\eea 
such that
\bea
 e_i^2 = 0 \ , \quad
 e_1 \cdot e_3 = e_1 \cdot e_4 = 0 \quad
 e_2 \cdot e_3 = e_2 \cdot e_4 = 0 \quad
 e_1 \cdot e_2 = - e_3 \cdot e_4 \ ,
\eea
and where $e_1$ and $e_2$ are real vectors, while $e_3$ and $e_4$ are complex. \\ 
The massless vectors $e_1$ and $e_2$ can be written as a linear combination of the two external
legs at the edges of the propagator carrying momentum ${\bar q}+p_0$, say $K_1$ and $K_2$,

\bea
 e_1^\nu = {1 \over \beta}\bigg(K_1^\nu + {K_1^2 \over \gamma} K_2^\nu \bigg) \ , \qquad
 e_2^\nu = {1 \over \beta}\bigg(K_2^\nu + {K_2^2 \over \gamma} K_1^\nu \bigg) \ ,  
\eea
with
\bea
\beta = 1 - {K_1^2 K_2^2 \over \gamma^2} \ , \quad {\rm and} \qquad
\gamma = K_1 \cdot K_2 
        + {\rm sgn}(1,K_1 \cdot K_2) \sqrt{ (K_1 \cdot K_2)^2 - K_1^2 K_2^2 } \ .
\eea
The massless vectors $e_3$ and $e_4$ can be then obtained as,
\bea
e_3^\nu = {\langle e_1| \gamma^\nu | e_2 ] \over 2} \ , \qquad 
e_4^\nu = {\langle e_2| \gamma^\nu | e_1 ] \over 2} \ .
\eea
In the case of double-cut, $K_1$ is the momentum flowing through the 
corresponding 2-point diagram, and $K_2$ is an arbitrary massless vector. 
In the case of single-cut, $K_1$ and $K_2$ cannot
be selected from the diagram, and are chosen as arbitrary vectors.

After defining the basis adopted for decomposing the solutions of the multiple-cuts, 
we can list the corresponding polynomial functions, whose variables are the 
components of the loop-momentum not-constrained by the cut-conditions.

\subsubsection{Quintuple Cut}

The residue of the quintuple-cut, 
$\db{i} = \ldots = \db{m} = 0$, defined as,
\bea
\Delta_{i j k \ell m}({\bar q}) = 
\Res_{i j k \ell m}\Bigg\{
{N({\bar q}) \over \db{0} \cdots \db{n-1}}
\Bigg\} 
\label{def:Resi5:left}
\eea
can be parametrized as \cite{Melnikov:2010iu},
\bea
\Delta_{i j k \ell m}({\bar q}) = c_{5,0}^{( i j k \ell m)} \ \mu^2 \ .
\label{def:Resi5:right}
\eea

\subsubsection{Quadruple Cut}

The residue of the quadruple-cut,  
$\db{i} = \ldots = \db{\ell} = 0$, defined as,
\bea
\Delta_{i j k \ell}({\bar q})= 
\Res_{i j k \ell}\Bigg\{
{N({\bar q}) \over \db{0} \cdots \db{n-1}}
- \sum_{i < \! < m}^{n-1}
         { \Delta_{ i j k \ell m}({\bar q}) \over 
           \db{i} \db{j} \db{k} \db{\ell} \db{m} } 
\Bigg\}  
\label{def:Resi4:left}
\eea
is parametrized as,
{\small \bea
 \Delta_{ i j k \ell}({\bar q}) 
\!\!&=&\!\! 
 c_{4,0}^{( i j k \ell)}
+c_{4,2}^{( i j k \ell)} \mu^2
+c_{4,4}^{( i j k \ell)} \mu^4 + 
\nn  
&& +\Big(
c_{4,1}^{( i j k \ell)}
+c_{4,3}^{( i j k \ell)} \ \mu^2
\Big)  
\Big[ (K_3 \cdot e_4) (q+p_0)\cdot e_3 
- (K_3 \cdot e_3) (q+p_0)\cdot e_4 \Big] = 
\qquad \nn
&& \nn
\!\!&=&\!\!
 c_{4,0}^{( i j k \ell)}
+c_{4,2}^{( i j k \ell)} \mu^2
+c_{4,4}^{( i j k \ell)} \mu^4 
-\Big(
c_{4,1}^{( i j k \ell)}
+c_{4,3}^{( i j k \ell)} \ \mu^2
\Big)  
\Big[ (K_3 \cdot e_4) x_4 - (K_3 \cdot e_3) x_3 \Big] (e_1 \cdot e_2) \ ,\nn
\label{def:Resi4:right}
\eea 
}
where $K_3$ is the third leg of the 4-point function associated to 
the considered quadruple-cut.

\subsubsection{Triple Cut}

The residue of the triple-cut, 
$\db{i} = \db{j} = \db{k} = 0$, defined as,
{\small \bea
\Delta_{i j k}({\bar q}) = 
\Res_{i j k}\Bigg\{
{N({\bar q}) \over \db{0} \cdots \db{n-1}}
- \sum_{i < \! < m}^{n-1}
         { \Delta_{ i j k \ell m}({\bar q}) \over 
           \db{i} \db{j} \db{k} \db{\ell} \db{m} } 
- \sum_{i < \! < \ell }^{n-1}
         { \Delta_{ i j k \ell}({\bar q}) \over 
           \db{i} \db{j} \db{k} \db{\ell}  } 
\Bigg\} 
\label{def:Resi3:left}
\eea }
is parametrized as,
{\small \bea
\Delta_{ i j k}({\bar q}) 
&=& 
 c_{3,0}^{( i j k)} +c_{3,7}^{( i j k)} \mu^2 + \nn 
&+&c_{3,1}^{( i j k)} (q+p_0)\cdot e_3
+c_{3,2}^{( i j k)} ((q+p_0)\cdot e_3)^2
+c_{3,3}^{( i j k)} ((q+p_0)\cdot e_3)^3 + \nn
&+&c_{3,4}^{( i j k)} (q+p_0)\cdot e_4
+c_{3,5}^{( i j k)} ((q+p_0)\cdot e_4)^2
+c_{3,6}^{( i j k)} ((q+p_0)\cdot e_4)^3 = \nn
&& \nn
&=& 
 c_{3,0}^{( i j k)} +c_{3,7}^{( i j k)} \mu^2 -
\Big(c_{3,1}^{( i j k)} x_4   + c_{3,4}^{( i j k)} x_3 \Big)   (e_1 \cdot e_2) + 
\nn 
&+&\Big(c_{3,2}^{( i j k)} x_4^2 + c_{3,5}^{( i j k)} x_3^2 \Big) (e_1 \cdot e_2)^2 -
\Big(c_{3,3}^{( i j k)} x_4^3 + c_{3,6}^{( i j k)} x_3^3 \Big) (e_1 \cdot e_2)^3 \ .
\qquad
\label{def:Resi3:right}
\eea}

\subsubsection{Double Cut}

The residue of the double-cut, 
$\db{i} = \db{j} = 0$, defined as,
{\small \bea
\Delta_{ i j}({\bar q}) 
=
\Res_{i j}\Bigg\{
{N({\bar q}) \over \db{0} \cdots \db{n-1}}
- \sum_{i < \! < m}^{n-1}
         { \Delta_{ i j k \ell m}({\bar q}) \over 
           \db{i} \db{j} \db{k} \db{\ell} \db{m} } 
- \sum_{i < \! < \ell }^{n-1}
         { \Delta_{ i j k \ell}({\bar q}) \over 
           \db{i} \db{j} \db{k} \db{\ell}  }  
- \sum_{i < \! < k }^{n-1}
         { \Delta_{ i j k }({\bar q}) \over 
           \db{i} \db{j} \db{k} } 
\Bigg\} \ , \qquad
\label{def:Resi2:left}
\eea }
can be interpolated by the following form,
{\small
\bea
\Delta_{ i j}({\bar q}) 
&=& 
 c_{2,0}^{(i j)} +c_{2,9}^{(i j)} \mu^2 + \nn 
&+&c_{2,1}^{(i j)} (q+p_0)\cdot e_2
+c_{2,2}^{(i j)} ((q+p_0)\cdot e_2)^2 + \nn
&+&c_{2,3}^{(i j)} (q+p_0)\cdot e_3
+c_{2,4}^{(i j)} ((q+p_0)\cdot e_3)^2 + \nn
&+&c_{2,5}^{(i j)} (q+p_0)\cdot e_4
+c_{2,6}^{(i j)} ((q+p_0)\cdot e_4)^2 + \nn
&+&c_{2,7}^{(i j)} ((q+p_0)\cdot e_2) ((q+p_0)\cdot e_3)
+c_{2,8}^{(i j)} ((q+p_0)\cdot e_2) ((q+p_0)\cdot e_4) 
= \nn
&& \nn
&=& 
  c_{2,0}^{(i j)} + c_{2,9}^{( i j )} \mu^2 +
\Big(
  c_{2,1}^{( i j )} x_1
- c_{2,3}^{( i j )} x_4 
- c_{2,5}^{( i j )} x_3
\Big)   (e_1 \cdot e_2) + \nn  
&+&\Big(
  c_{2,2}^{( i j )} x_1^2 
+ c_{2,4}^{( i j )} x_4^2 
+ c_{2,6}^{( i j )} x_3^2 
- c_{2,7}^{( i j )} x_1 x_4  
- c_{2,8}^{( i j )} x_1 x_3
\Big)  (e_1 \cdot e_2)^2 \ .
\label{def:Resi2:right}
\eea }

\subsubsection{Single Cut}
The residue of the single-cut, $\db{i} = 0$, defined as,
\bea
\Delta_{ i }({\bar q}) 
&=& 
\Res_{i}\Bigg\{
{N({\bar q}) \over \db{0} \cdots \db{n-1}}
- \sum_{i < \! < m}^{n-1}
         { \Delta_{ i j k \ell m}({\bar q}) \over 
           \db{i} \db{j} \db{k} \db{\ell} \db{m} } 
- \sum_{i < \! < \ell }^{n-1}
         { \Delta_{ i j k \ell}({\bar q}) \over 
           \db{i} \db{j} \db{k} \db{\ell}  }  + \nn && \qquad \quad
- \sum_{i < \! < k }^{n-1}
         { \Delta_{ i j k }({\bar q}) \over 
           \db{i} \db{j} \db{k} } 
- \sum_{i < j  }^{n-1}
         { \Delta_{ i j }({\bar q}) \over 
           \db{i} \db{j} } 
\Bigg\} 
\label{def:Resi1:left}
\eea
can be interpolated as follows,
\bea
\Delta_{ i }({\bar q}) 
&=& 
  c_{1,0}^{(i)} 
+ c_{1,1}^{(i)} ((q+p_0)\cdot e_1)
+ c_{1,2}^{(i)} ((q+p_0)\cdot e_2) + \nn
&+& c_{1,3}^{(i)} ((q+p_0)\cdot e_3)
+ c_{1,4}^{(i)} ((q+p_0)\cdot e_4)
= \nn
&& \nn
&=& 
  c_{1,0}^{(i)} + 
\Big(
  c_{1,1}^{(i)} x_2 
+ c_{1,2}^{(i)} x_1
- c_{1,3}^{(i)} x_4 
- c_{1,4}^{(i)} x_3
\Big) (e_1 \cdot e_2) \ .
\label{def:Resi1:right}
\eea

\subsubsection{Discrete Fourier Transform}

As proposed in \cite{Mastrolia:2008jb},
the coefficients of a polynomial of degree $n$ in the variable $x$, say $P(x)$, defined as,
\bea
P(x) = \sum_{\ell=0}^n \ c_\ell \ x^\ell \ ,
\label{eq:GeneralPn}
\eea
can be extracted by means of {\it projections}, according to the 
the Discrete Fourier Transform. The basic procedure is very simple:

\begin{enumerate}
\item
generate the set of discrete values $P_k \ (k=0,...,n)$,
\bea
P_{k} = P(x_k) = 
\sum_{\ell=0}^n \ c_\ell \ \rho^\ell \ e^{-2 \pi i {k \over (n+1)} \ell} \ ,
\eea
by sampling $P(x)$ at the points
\bea
 x_k = \rho \ e^{-2 \pi i {k \over (n+1)}} \ ;
\eea
\item
using the orthogonality relation
\bea
\sum_{n=0}^{N-1} e^{2 \pi i {k \over N} \ n} \ e^{-2 \pi i{k^\prime \over N} \ n}
 = N \ \delta_{k k'} \ ,
\eea
each coefficient $c_\ell$ finally reads,
\bea
c_\ell &=& {\rho^{-\ell} \over n+1}
         \sum_{k=0}^n \ P_{k}  \  e^{2 \pi i {k \over (n+1)} \ell} \ .
\eea
\end{enumerate}
The extension of the DFT projection to the case of multi-variate polynomials
is straightforward. \\
As one can notice the formula for the coefficients $c_\ell$, although simple,
diverges when $\rho$ goes to zero. By using the parametrization
in Eq.(\ref{eq:q_deco}), the radius
$\rho$ happens to be constrained by the on-shell cut-condition.  
Depending on the external invariants and internal masses, the dangerous 
value $\rho=0$ might occur. 
In a previous work \cite{Mastrolia:2008jb}, we described a 
safer sampling, which significantly reduces the numerical instabilities arising
from the vanishing of $\rho$. We do not repeat the same discussion here,
but recall that the sampling of the multiple-cut solutions used for the polynomial
interpolation of the triple- and double-cut residues within {\samurai} are chosen according to that algorithm.
By using the DFT solutions as described in \cite{Mastrolia:2008jb},
we sample the numerator functions exactly as many times as the number 
of the unknown coefficients, without needing additional sampling points 
to improve the numeric precision, which would demand more computing time.

\subsection{Amplitude and Master Integrals}
\label{sec:AmpAndMI}

The knowledge of all the coefficients appearing in the polynomials
$\Delta_{ i j k \ell m}$, $\Delta_{ i j k \ell}$, $\Delta_{ i j k}$, $\Delta_{ i j}$,  
and $\Delta_{ i }$ implies the following expression for the one-loop $n$-point amplitude,
\bea
{\cal A}_n &=& 
\sum_{i < j < k < \ell}^{n-1}\bigg\{
          c_{4,0}^{ (i j k \ell)} I_{i j k \ell}^{(d)} 
-{(d-4) \over 2} 
          c_{4,2}^{ (i j k \ell)} I_{i j k \ell}^{(d+2)} 
+{(d-2)(d-4) \over 4}
          c_{4,4}^{ (i j k \ell)} I_{i j k \ell}^{(d+4)} 
\bigg\} \nn
     &+&
\sum_{i < j < k }^{n-1}\bigg\{
          c_{3,0}^{ (i j k)} I_{i j k}^{(d)} 
-{(d-4) \over 2} 
          c_{3,7}^{ (i j k)} I_{i j k}^{(d+2)} 
\bigg\} \nn
     &+&
\sum_{i < j }^{n-1}\bigg\{
          c_{2,0}^{ (i j)} I_{i j}^{(d)} 
        + c_{2,1}^{ (i j)} J_{i j}^{(d)} 
        + c_{2,2}^{ (i j)} K_{i j}^{(d)} 
-{(d-4) \over 2} 
          c_{2,9}^{ (i j)} I_{i j}^{(d+2)} 
\bigg\} \nn
     &+&
\sum_{i}^{n-1}
          c_{1,0}^{ (i)} I_{i}^{(d)} 
 \ ,
\label{eq:Aresult}
\eea
where, beside the scalar boxes, triangles, bubbles and tadpoles, 
the other master integrals are \cite{Pittau:1996ez,Bern:1995db}
\bea
\int d^d {\bar q} {\mu^2 \over \db{i} \db{j} \db{k} \db{\ell} } 
&=& -{(d-4) \over 2} I_{i j k \ell}^{(d+2)} \ , \\
\int d^d {\bar q} {\mu^4 \over \db{i} \db{j} \db{k} \db{\ell} } 
&=& {(d-2) (d-4) \over 4} I_{i j k \ell}^{(d+4)} \ , \\
\int d^d {\bar q} {\mu^2 \over \db{i} \db{j} \db{k} } 
&=& -{(d-4) \over 2} I_{i j k}^{(d+2)} \ , \\ 
\int d^d {\bar q} {\mu^2 \over \db{i} \db{j} } 
&=& -{(d-4) \over 2} I_{i j}^{(d+2)} \ , \\
\int d^d {\bar q} {{\bar q} \cdot e_2 \over \db{i} \db{j} } 
&=&  J_{i j}^{(d)} \ , \\
\int d^d {\bar q} {({\bar q} \cdot e_2)^2 \over \db{i} \db{j} } 
&=&  K_{i j}^{(d)} \ .
\eea
The last two master integrals, $J_{i j}^{(d)}$ and $K_{i j}^{(d)}$, respectively
a linear and a quadratic 2-point function, appear as a consequence of the polynomial structure
of $\Delta_{ i j }({\bar q})$, defined in Eq.({\ref{def:Resi2:right}}), which was chosen to have 
no singularity in presence of vanishing external invariant~\cite{Ossola:2007bb}. The vector
$e_2$ entering their definition is an element of the loop-momentum basis, defined in Eq.(\ref{eq:q_deco}), 
and used for the solutions of the double-cut $\db{i} = \db{j}=0$.
Also, because of the monomial parametrization of the quintuple-cut residue,
$\Delta_{ i j k \ell m}({\bar q})$, given in Eq.({\ref{def:Resi5:right}}),
the decomposition of the amplitude in terms 
of MI, Eq.(\ref{eq:Aresult}), is free of scalar pentagons, as already
noticed in \cite{Melnikov:2010iu}.

\section{Running {\samurai} }

In this section we give some details about using {\samurai}.
All the files are available on the webpage:
\begin{center}
{\tt http://cern.ch/samurai}
\end{center}

\noindent
The archive {\tt samurai\_v1.0.tar.gz} contains the files for the {\samurai} library, several examples of calculations, and also the routines 
for the evaluation scalar integrals QCDLoop~\cite{Ellis:2007qk} and OneLOop~\cite{vanHameren:2009dr}.

\begin{enumerate}
\item Download the archive {\tt samurai\_v1.0.tar.gz} and extract the files. They will be copied in a folder
called {\tt \slash samurai}.
\item Run the {\tt Install} script. It will compile all useful routines and organize them. 
All routines are written in Fortran~90 and the default compiler is {\tt gfortran}. In order to change compiler (or compiling options), the user should edit all the {\tt makefile} commands.

After running the {\tt Install} script, you will find four subfolders within the {\tt \slash samurai} directory:
the subdirectory named {\tt \slash libs} will contain all the libraries, namely the reduction routines {\tt libsamurai.a}, and three libraries for the 
numerical evaluation of the master integrals.

Examples that reproduce all calculations described in Sec. \ref{sec:Examples}
can be found in separate subfolders in {\tt \slash examples}.
The {\tt Install} script compiles all the examples, with the exception of the ``Six Quarks'' (that takes about
10 minutes to compile). The user can process it separately by typing {\tt make} in the directory {\tt \slash examples\slash uussbb}.

\item Run each process using the corresponding command {\tt process.exe}. 
\end{enumerate}

\noindent
The use of {\samurai} is implemented through the following
chain of calls is:

{\tt call initsamurai(imeth,isca,verbosity,itest)} 

{\tt call InitDenominators(nleg,Pi,msq,v0,m0,v1,m1,...,vlast,mlast)} 

{\tt call samurai(xnum,tot,totr,Pi,msq,nleg,rank,istop,scale2,ok)} 

{\tt call exitsamurai}

\subsection{Initialization}

To initialize the {\samurai} library, one needs to choose
the arguments of the subroutine {\tt initsamurai}

\begin{center}
{\tt call initsamurai(imeth,isca,verbosity,itest)}
\end{center}

\noindent
which specify the 
the type of input to reduce ({\tt imeth}),
the routines for the numerical evaluation of the scalar integrals ({\tt isca}),
the details of the output ({\tt verbosity}), and the test to apply 
to the reconstruction ({\tt itest}):


\begin{itemize}
\item{\tt imeth -}
  {\samurai} can reduce integrands of one-loop amplitudes 
defined either as {\it numerator of diagrams} sitting on products 
of denominators, specified with {\tt imeth=diag};
or as {\it products of tree-level amplitudes} sewn along cut-lines,
specified with {\tt imeth=tree}.

\item{\tt isca -}
 The user can trigger the use of QCDLoop~\cite{Ellis:2007qk}
by assigning {\tt isca=1},
or the use of OneLOop~\cite{vanHameren:2009dr} with {\tt isca=2}.

\item{\tt verbosity -}
 The level of information printed in the file {\tt output.dat}
can be chosen with the value of {\tt verbosity:}

 {\tt verbosity=0}, no output;

 {\tt verbosity=1}, the coefficients are printed;

 {\tt verbosity=2}, the value of the MI's are printed as well;

 {\tt verbosity=3}, the outcome of the numerical test appears.

\item{\tt itest -} 
 This option is used to select the test to monitoring the quality
of the numerical reconstruction.
The possibilities are: {\tt itest=0,1,2,3} to have respectively
none, the global $(N=N)$-test, the local $(N=N)$-test, and the power-test,
which are described in Sec.\ref{sec:tests}

While {\tt imeth=diag} supports all the options for {\tt itest},
the choice {\tt imeth=tree} allows only {\tt itest=0,2}.

\end{itemize}

\subsection{Integrand definition}

After selecting the routines for the scalar integrals and the reduction technique, the user should 
provide information about the integrand, by specifying the {\it numerator} and 
the {\it denominators}.

\bigskip

The {\it denominators} of the diagram to be reduced are defined through 
the subroutine {\tt InitDenominators} 
which generates the lists of internal momenta {\tt Pi} and squared masses 
{\tt msq} characterizing each propagator:

\begin{center}
{\tt call InitDenominators(nleg,Pi,msq,v0,m0,v1,m1,...,vlast,mlast)}
\end{center}

\noindent
The arguments of the subroutine, 
labeled as input/output ({\tt [i/o]}) according to their role, are:

\begin{itemize}
\item{\tt nleg -} 
{\tt [i]}. The integer number of the external legs of the diagram, 
corresponding to the number of denominators.
\item{\tt Pi -} 
{\tt [o]}. The array {\tt Pi(i,m)} contains the {\tt nleg} four-vectors present in the denominators of the integrand, namely the vectors $p_i$ of Eq.(\ref{def:An}) where we used the definition $\db{i} = ({\bar q} + p_i)^2-m_i^2$. 
In the notation {\tt Pi(i,m)}, the first index, 
{\tt i=0,\dots,nleg-1}, 
runs on the set of the denominators; while the second index 
{\tt m=1,\dots,4}, runs over the components of the vector, with the energy being given
as $4^{\rm th}$ component.
\item{\tt msq -} 
{\tt [o]}. The array {\tt msq(i)}, is the list of the squared masses that appear 
in the propagators. The ordering {\tt i=0,\dots,nleg-1} is bound to
the list of momenta {\tt Pi(i,m)}.
\item{\tt v0, m0 -} 
{\tt [i]}. The vector {\tt v0} and the mass {\tt m0} are assigned 
to the first denominator. 
\item{\tt vlast, mlast -} 
{\tt [i]}. The vector {\tt vlast} and the mass {\tt mlast} are assigned to the last denominator. 
\end{itemize}

\subsection{Reduction}

Having defined the integrand denominators, 
characterized by {\tt Pi} and {\tt msq}, the actual reduction of the 
input ({\tt xnum}) is performed by the library {\samurai},

\begin{center}
{\tt call samurai(xnum,tot,totr,Pi,msq,nleg,rank,istop,scale2,ok)}
\end{center}

\noindent
which writes the total result of the reduction in {\tt tot}. 
For convenience, the rational term 
is also separately written in {\tt totr}.

Here comes the detailed description of each argument:
\begin{itemize}
\item{\tt xnum -} 
{\tt [i]}. 
The {\it numerator} of the diagram is defined in an external function, 
whose name can be decided by the user, but with fixed arguments.
Hereby we adopt the dummy name {\tt xnum}.

The complex function {\tt xnum(icut,q,mu2)} is the
integrand to be reduced. 
The arguments of the function {\tt xnum(icut,q,mu2)} are: 

{\tt icut}, an integer labeling the cut, where each digit 
corresponds to a cut-denominator in descending order 
(ex. {\tt icut}$=3210$ corresponds to
the quadruple-cut $\db{0}=\db{1}=\db{2}=\db{3}=0$ );

{\tt q}, the virtual four-momentum, $q$ (with the energy given as $4^{\rm th}$ component);

and {\tt mu2} the extra-dimensional mass-scale, $\mu^2$.

When {\tt imeth=diag}, {\tt xnum} is expected to have the form 
of a numerator, hence being polynomial in $q$ and $\mu^2$.
In this case {\tt xnum} is a unique function to be processed 
at every level of the top-down reduction by cycling on {\tt icut}, 
but does not depend on the considered cut.

When {\tt imeth=tree}, {\tt xnum} is expected to be formed by the product of tree-amplitudes, therefore the presence of propagators it is also allowed.
In this case, {\tt xnum} is not unique, but should change according to 
the considered cut. Therefore, the value of {\tt icut} yields a selective access 
to the proper integrand within the same function.

\item{\tt tot -}
{\tt [o]}. The complex variable {\tt tot} contains the final result for the integrated amplitude of numerator {\tt xnum}. The finite part, that also includes the rational term, will be stored in {\tt tot(0)}, while {\tt tot(-1)} and 
{\tt tot(-2)} contain the single and double poles, respectively.
\item{\tt totr -}
{\tt [o]}. For the purpose of comparisons and debugging, we also provide the rational part {\tt totr} alone.
This complex number is the sum of all contributions coming from integrals in shifted dimensions, namely all contributions that contain 
a dependence from $\mu^2$ in the reconstructed integrand.
\item{\tt nleg -}
{\tt [i]}. Already defined.
\item{\tt Pi -}
{\tt [i]}. Already defined.
\item{\tt msq -} 
{\tt [i]}. Already defined.
\item{\tt rank -}
{\tt [i]}. This integer value is the maximum rank of the numerator. This information is extremely valuable in order
to optimize the reduction and improve the stability of the results. Using this information, we can simplify the reconstruction of the numerator by eliminating contributions that do not appear in the reduction.
If the information about the rank of the integrand is not available, {\tt rank} should be set equal to {\tt nleg}. 
\item{\tt istop -}
{\tt [i]}. This flag stops the reduction at 
the level requested by the user. 
{\tt istop} is an integer, whose range of values is from 1 to 5.

{\tt istop=5,4,3,2,1} will interrupt the calculation 
after determining pentagon, box, triangle, bubble, 
and tadpole coefficients respectively.
This procedure can be particularly useful to improve the precision 
of calculations when one knows a priori that a particular set of 
integrals does not contribute to the considered process. 
\item{\tt scale2 -}
{\tt [i]}. This is the scale (squared) that is used in the evaluation of scalar integrals.
\item{\tt ok -} 
{\tt [o]}. This logical variable carries information about the goodness of the
reconstruction. The default values is {\tt ok=true}, and it is set 
to {\tt ok=false} when the reconstrucion test fails.
\end{itemize}

\bigskip

As stated in Section \ref{sec:numerator}, the generic one-loop integrand 
can be polynomial in $\epsilon$ up to the second-order.
Each coefficient of the $\epsilon$-decomposition can be assigned to a specific
function, {\it i.e.} {\tt xnum0}, {\tt xnum1}, {\tt xnum2}, 
which can be independently processed.

\subsection{Reconstruction Tests}
\label{sec:tests}

There are three different ways of monitoring the quality of the coefficients 
reconstructed by {\samurai}.

\subsubsection{Global $(N=N)$-test}

The first option ({\tt itest=1}) is the so-called ``$N=N$'' test on the reconstructed 
expression for the numerator functions, which was already discussed 
in \cite{Ossola:2006us,Ossola:2007bb}. 
It is based on the equality given by Eq.(\ref{def:OPP:deco}),
between the original numerator in the {\it l.h.s.} and the reconstructed one 
in the {\it r.h.s.}, evaluated at an arbitrary value of ${\bar q}$. \\
A possible drawback of this precision test lies in the fact that the coefficients of tadpoles and bubbles in Eq.(\ref{def:OPP:deco}) multiply
a large set of denominators: for a six-point function, each tadpole coefficient multiplies five denominators, namely a term proportional to masses 
or momenta, $q$, raised to ten powers, that can be huge in some cases or very small in other situations.
 This might have the effect of hiding the contribution of some coefficients or, as happens more frequently, might yield to overestimating the error in the reconstruction.

\subsubsection{Local $(N=N)$-test}

A second check is a ``local $N=N$'' test ({\tt itest=2}), 
regarding the reconstruction of each polynomial $\Delta({\bar q})$,
respectively defined in Eqs.(\ref{def:Resi5:left}, \ref{def:Resi4:left}, \ref{def:Resi3:left}, \ref{def:Resi2:left}, \ref{def:Resi1:left}).
In this case the value of ${\bar q}$ used for the numerical check 
is chosen among additional solutions of the considered multiple-cut, which have 
not participated to the determination of $\Delta({\bar q})$ itself.
This option is suitable for a unitarity-based calculation ({\tt imeth=tree}).

\subsubsection{Power-test}

A third option ({\tt itest=3}) for testing the precision of the 
reconstruction is the ``power test''.
We can observe that the maximum powers in $q$ in the {\it r.h.s} and {\it l.h.s} of Eq.(\ref{def:OPP:deco}) are different: 
the reconstructed side can contain terms with high powers of $q$ 
that are not present in the original numerator.
Therefore it is clear that the overall coefficients 
in front of these terms should vanish. \\
The reconstructed expressions in general are not simple, since they involve pieces coming from the polynomial spurious terms multiplied by the denominators. However, \emph{for each choice of the rank and number of denominators, there is at least one simple set of coefficients that sum to zero} exactly. Moreover, this set is the lowest one in the reconstruction and therefore it carries information about any loss of precision at previous steps in the reduction.\\
If the difference between the rank and the number of denominator is equal to three
({\tt nleg-irank} {\tt=3}), the sum of all the coefficients of three-point scalar integrals should be zero, namely:
\begin{equation}
\sum_i c_{3,i}(0) = 0
\end{equation}
where the sum is over all possible triple cuts. \\
Analogously, if the difference between the rank and number of denominator is equal to two ({\tt nleg-irank=2}), the sum of the coefficients of two-point scalar integrals should be zero, namely:
\begin{equation}
\sum_i c_{2,i}(0) = 0
\end{equation}
where the sum involves all double cuts.\\
Finally, if the difference between the rank and number of denominator is equal to one,
({\tt nleg-irank=1}), 
the sum of the coefficients of the tadpole scalar integrals should be zero, namely:
\begin{equation}
\sum_i c_{1,i}(0) = 0
\end{equation}
where the sum involves all single cuts.\\
The situation is slightly more complicated for maximum rank when difference between the rank and number of denominator is equal to zero. 
If ({\tt nleg-irank=0}), we should consider all the one-point spurious coefficients $c_{1,i}(1)$ to $c_{1,i}(4)$, each multiplied by the corresponding vector $e_{1,i}, \ldots, e_{4,i}$ of the basis defined in Section \ref{sec:Polynomials}.
Summing over all possible single cuts, labeled by ${i}$, we get the condition
\begin{equation}
\sum_i \sum_{n=1}^{4} c_{1,i}(n) e_{n,i}^\mu = 0
\end{equation}
As a final remark, we observe that the outcome of the ``power test'' does not depend in any way from the choice of the integrated momentum ${\bar q}$, 
unlike the previous two methods.

\bigskip
\noindent
The threshold values for the reconstruction checks can be set 
in the file {\tt ltest.dat}, to be located in the directory where 
the call to {\tt initsamurai} is made. 
The phase-space points failing the tests ({\tt ok=false}) are 
stored in the file {\tt bad.points}, in the same directory.
In principle they could be re-processed enhancing the numerical precision
by compiling the {\samurai} library in quadruple-precision.

\subsection{Comments on Precision}
\label{sec:Precision}

The precision of the results obtained using a reduction algorithm
at the integrand-level depends on many variables. 

When the numerator is a real function of the
external momenta and masses there is a simple way to establish the quality
of the reduction: real functions give
rise to real coefficients of MI. 
In this case, the error on each coefficient can be estimated by the size of the 
imaginary part, that should vanish. 

More generally,
the quality of the reconstruction can be quantified 
by the ratio of 
the difference between the
exact calculation~(analytical or multi-precision) and the
reconstructed one, and the former, evaluated over a large
set of unweighted points. 
This procedure gives a good indication, but it is not always safe, 
because the error on the
prediction in a calculation based on the importance sampling
could suffer from the accumulation of bad points in the neighborhood of 
higher weights.

We identify three kinds of possible instabilities, which could be 
all controlled by adopting quadrupole or multiple precision routines. 

\noindent
The first kind of instabilities is related to the well known problem 
of the vanishing of the Gram determinants, 
inducing an enhancement of the coefficients 
of the MI carrying such pathological kinematic factor.
They can be monitored by the tests
implemented in \samurai, and the dangerous cases could be dealt with 
by introducing branches to dedicated reduction routines,
hence without making use of the multiple precision.

\noindent
The second kind corresponds to big cancellations among 
the contributions from different diagrams in the same calculation.
On-shell methods, which work with purely gauge invariant objects,
seems to represent the best option to avoid such problem.

\noindent
The third type of instability can occur when the values of 
internal masses are sensibly larger then the phase-space invariants.
In this case, both the cut-constructible part and the
rational term are large but their sum remains relatively small.
This in principle could be cured with a change of the integral basis
where the cancellations are built-in.

Our tool does not switch automatically between double and quadruple precision.
The running in the latter case is time-consuming, 
therefore, along the lines of the above considerations, 
we are investigating a more systematic treatment 
of the problematic configurations, which
goes beyond the scope of this version of the code, 
and will be the subject of a future publication.

\section{Examples of Applications}
\label{sec:Examples}

In this section we present examples of calculations of
one-loop amplitudes performed with {\samurai}.
These examples are selected with the idea of covering different
situations and problems that can be treated within our code.
Our intention is to show the flexibility of this framework and present 
examples of applications performed in various regularization
schemes widely used for the calculation of one loop virtual corrections.

{\samurai} can process two different kinds of input,
according to the strategy adopted for the generation of the
integrand. In the Feynman diagrams approach one should provide
a set of numerator functions, each accompanied by a corresponding
list of denominators. On the other hand, in the generalized
unitarity approach the input will be in the form of products
of tree-level amplitudes. In the following we describe some
calculations performed within both frameworks.

In several cases, we use Rambo \cite{Kleiss:1985gy} for generating
phase-space points.

\subsection{Four-photon Amplitudes}

This example is useful to verify the proper reconstruction 
of the rational term.
The leading term of the process $\gamma \gamma \to \gamma \gamma$ in QED 
proceeds {\it via} fermion-loop~\cite{Gounaris:1999gh,Bernicot:2008th}.
We treat both cases of massless and massive fermion.
The four-photon amplitudes get contributions from the 6 Feynman diagrams
representing the possible permutations of the 4 photons attached to the
fermion loop. Indeed, only 3 permutations are independent and need
to be evaluated, because loops related by flipping the fermion line
give the same answer.
Let us consider the diagram with the photons labeled in clockwise order 1234, carrying the following denominators,
\bea
({\bar L}_1^2-m^2) \ 
({\bar L}_2^2-m^2) \ 
({\bar L}_3^2-m^2) \ 
({\bar L}_4^2-m^2)
\eea
and {\it numerator}, 
\bea
N({\bar q}) = -{\rm Tr} \Big[ 
({\bar {\slh L}}_1+m)\,\slh \epsilon_2\,
({\bar {\slh L}}_2+m)\,\slh \epsilon_3\,
({\bar {\slh L}}_3+m)\,\slh \epsilon_4\,
({\bar {\slh L}}_{4}+m)\,\slh \epsilon_1 \Big] 
\eea
where
\bea
{\bar L}_1 = {\bar q} \ , \ 
{\bar L}_2 = {\bar q} + p_2 \ , \ 
{\bar L}_3 = {\bar q} + p_{23} \ , \ 
{\bar L}_4 = {\bar q} + p_{234} \ , \ 
\eea
with $p_{ijk} = p_i+p_j+p_k$.
The other two independent contributions are obtained by
permuting momenta and polarizations: $(234) \rightarrow (243),~(324)$.
So, in this example, the inputs to run {\samurai} are simply:
${\tt Pi}=(0,p_2,p_2+p_3,p_2+p_3+p_4),~{\tt msq}=(m^2,m^2,m^2,m^2),~
{\tt irank}=4,~{\tt istop}=1$.
Once the loop momentum is decomposed as in Eq.(\ref{eq:qmu}), 
we end up with an expression suitable for the numerical evaluation:
\bea \label{eqabove}
N(q, \mu^2) &=& - ( m^4 - \mu^2\,m^2 + \mu^4 )\,{\rm Tr}[
\slh \epsilon_2\,\slh \epsilon_3\,\slh \epsilon_4\,\slh \epsilon_1 ] \nn
&-& ( m^2 -  \mu^2 )  
\Big( {\rm Tr}[\slh \epsilon_2 \, \slh \epsilon_3 \, \slh \epsilon_4 \, \slh L_4 \, \slh \epsilon_1 \, \slh   L_1] 
   +\,{\rm Tr}[\slh \epsilon_2 \, \slh \epsilon_3 \, \slh  L_3 \, \slh \epsilon_4 \, \slh \epsilon_1 \, \slh  L_1] \nn
&+& \,{\rm Tr}[\slh \epsilon_2 \, \slh \epsilon_3 \, \slh  L_3 \, \slh \epsilon_4 \, \slh L_4 \, \slh \epsilon_1] 
 +  \,{\rm Tr}[\slh \epsilon_2 \, \slh L_2 \, \slh \epsilon_3 \, \slh \epsilon_4 \, \slh \epsilon_1 \, \slh L_1] \nn
 &+&\,{\rm Tr}[\slh \epsilon_2 \, \slh L_2 \, \slh \epsilon_3 \, \slh \epsilon_4 \, \slh L_4 \, \slh \epsilon_1] 
   +\,{\rm Tr}[\slh \epsilon_2 \, \slh L_2 \, \slh \epsilon_3 \, \slh  L_3 \, \slh \epsilon_4 \, \slh \epsilon_1]\,\Big) \nn
   &-&{\rm Tr}[\slh  L_1 \, \slh \epsilon_2\, \slh L_2\, \slh \epsilon_3\, \slh L_3 \, \slh \epsilon_4\, \slh L_4\, \slh \epsilon_1 ] \ ,
\eea
where $L_i $ is the 4-dimensional part of ${\bar L}_i \ (= L_1 + \mu)$. 
Note that now the whole expression can be evaluated numerically in terms of
the four dimensional complex variable $q$ and 
the real variable $\mu^2$.
Using {\samurai}, it is easy to see that: the term proportional to $\mu^2\,m^2$ 
in Eq.(\ref{eqabove}) gives rise to null integrals and does not contribute, 
the terms proportional to $\mu^2q^\mu\,q^\nu$ are not individually zero but they cancel 
when summing over all contributions; and finally that the $\mu^4$-term gives
the correct rational term.

\subsection{Six-photon Amplitudes}

The six-photon amplitudes~\cite{Mahlon:1993fe,Nagy:2006xy,Binoth:2007ca,Ossola:2007bb,Gong:2008ww,Bernicot:2007hs,Bernicot:2008nd}
are also a good test for the reconstruction of the rational term,
that, after summing over all diagrams, has to vanish \cite{Binoth:2006hk}. \\ 
The construction of the amplitudes 
follow closely the one that we used for the four photons. 
Out of the 120 contributing diagrams, all containing up to rank-6 
tensor integrals, only 60 need to be computed. 
We can construct all of them as permutation of 
just one diagram. 
In the massless case, 
we consider the diagram with the photons in the clockwise order 123456, whose corresponding 
numerator reads,
\beq
N(q, \mu^2) = -{\rm Tr} \Big[ 
{\bar {\slh L}}_1\, \slh \epsilon_2\,
{\bar {\slh L}}_2\, \slh \epsilon_3\,
{\bar {\slh L}}_3\, \slh \epsilon_4\,
{\bar {\slh L}}_4\, \slh \epsilon_5\,
{\bar {\slh L}}_5\, \slh \epsilon_6\,
{\bar {\slh L}}_6\, \slh \epsilon_1 \Big]. 
\eeq
where
\bea
&&{\bar L}_1 = {\bar q} \ , \ 
{\bar L}_2 = {\bar q} + p_2 \ , \ 
{\bar L}_3 = {\bar q} + p_{23} \ , \ 
{\bar L}_4 = {\bar q} + p_{234} \ , \ \nn
&&
{\bar L}_5 = {\bar q} + p_{2345} \ , \ 
{\bar L}_6 = {\bar q} + p_{23456} \ .
\eea
This example turns out to be challenging for the reduction
algorithm, because each diagram separately admits a non-trivial
reduction with non-vanishing coefficients for all the MI and
rational terms but, after summing together the partial results of all
diagrams, there are strong cancellations. In the final answer 
all contributions coming from 2-point functions cancel out. 
Moreover, also the rational terms vanish. \\
Indeed, the final expression contains only cut-constructible terms
and no rational part and the knowledge of the coefficients of
boxes and triangles alone is sufficient to obtain the correct answer
for the total amplitude. 

After the dimensional decomposition of the loop momentum ${\bar q}$, 
it is easy to see that all the terms containing one, two or
three powers of $\mu^2$ give rise to vanishing integrals
and do not contribute.
As a consequence, the only term needed in the numerical evaluation is
the four dimensional one:
\beq
N(q, \mu^2) = N(q) = -{\rm Tr} \Big[ 
{ {\slh L}}_1\, \slh \epsilon_2\,
{ {\slh L}}_2\, \slh \epsilon_3\,
{ {\slh L}}_3\, \slh \epsilon_4\,
{ {\slh L}}_4\, \slh \epsilon_5\,
{ {\slh L}}_5\, \slh \epsilon_6\,
{ {\slh L}}_6\, \slh \epsilon_1 \Big]. 
\eeq
For a numerical check we consider the value of the
two amplitudes $A(-,-,+,$ $+,+,+)$ and $A(+,-,-,+,+,-)$ \cite{Bernicot:2007hs,Bernicot:2008nd}
\bea
\frac{s}{\alpha^3}\,A(-,-,+,+,+,+)&=&  11075.04009210435 \ ,\\
\frac{s}{\alpha^3}\,A(+,-,-,+,+,-)&=&  7814.762085902767 \ ,
\eea
evaluated at the phase-space point \cite{Nagy:2006xy}, 
\bea
{\vec p}_3 &=& ( 33.5, 15.9, 25.0) \\
{\vec p}_4 &=& (-12.5, 15.3,  0.3) \\
{\vec p}_5 &=& (-10.0,-18.0,- 3.3) \\
{\vec p}_6 &=& (-11.0,-13.2,-22.0)
\label{eq:6photonpsp}
\eea
with $p_1$ and $p_2$ directed along the positive and negative
$z$-axis respectively.

By running {\samurai} with ${\tt istop}=2$, namely keeping
the contributions of the bubbles, the results are:
\bea
\frac{s}{\alpha^3}\,A(-,-,+,+,+,+)&=&\underline{11075.040}174990 \ , \\
\frac{s}{\alpha^3}\,A(+,-,-,+,+,-)&=&\underline{7814.762}3429908 \ .
\eea
By exploiting the knowledge that contributions from bubbles 
and rational terms will vanish, and therefore removing these terms from the
reduction, we verify an improvement on the final result.
Infact, by setting ${\tt istop}=3$ and isolating only the cut-constructible 
terms (by subtracting {\tt totr} diagram by diagram),
the results of {\samurai} turn out to be in better agreement:
\bea
\frac{s}{\alpha^3}\,A(-,-,+,+,+,+)&=&\underline{11075.04009210}2 \ , \\
\frac{s}{\alpha^3}\,A(+,-,-,+,+,-)&=&\underline{7814.76208590}84 \ .
\eea

As expected, the strong cancellations between the 60 diagrams spoil
the precision of the full results even if the number of good digits
for this specific phase-space point can still be considered sufficient 
for phenomenological studies.

\subsection{Eight-photon Amplitudes}

\begin{figure}
\begin{center}
\includegraphics[width=0.8\textwidth]{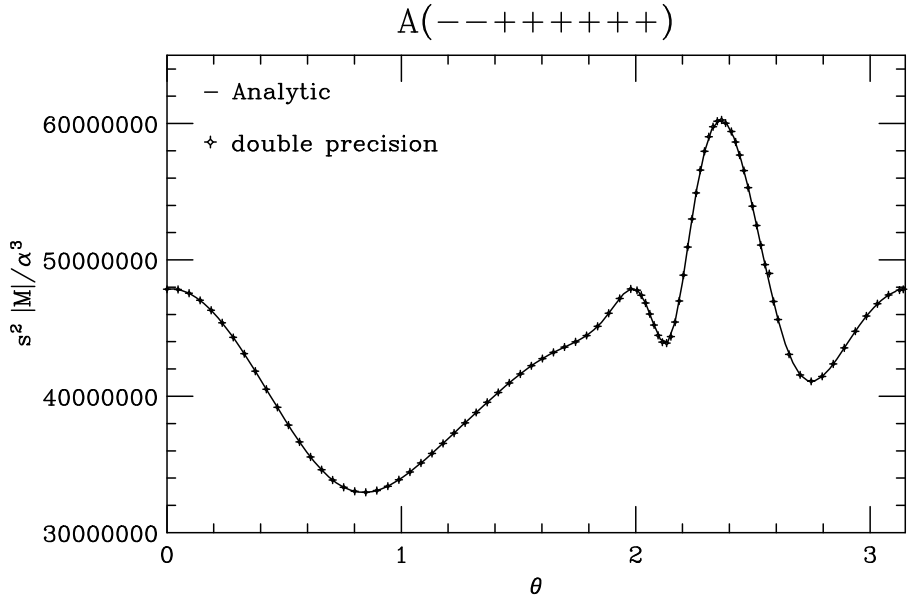}
\end{center}
\caption{
Results for the 8-photon amplitude with helicity $--++++++$.
The continuous line represents the analytic results of \cite{Mahlon:1993fe}.
The results of {\samurai} are produced in double-precision and with {\tt istop=4}.
}\label{fig:8photnoMHV}
\end{figure}

\begin{figure}
\begin{center}
\includegraphics[width=0.8\textwidth]{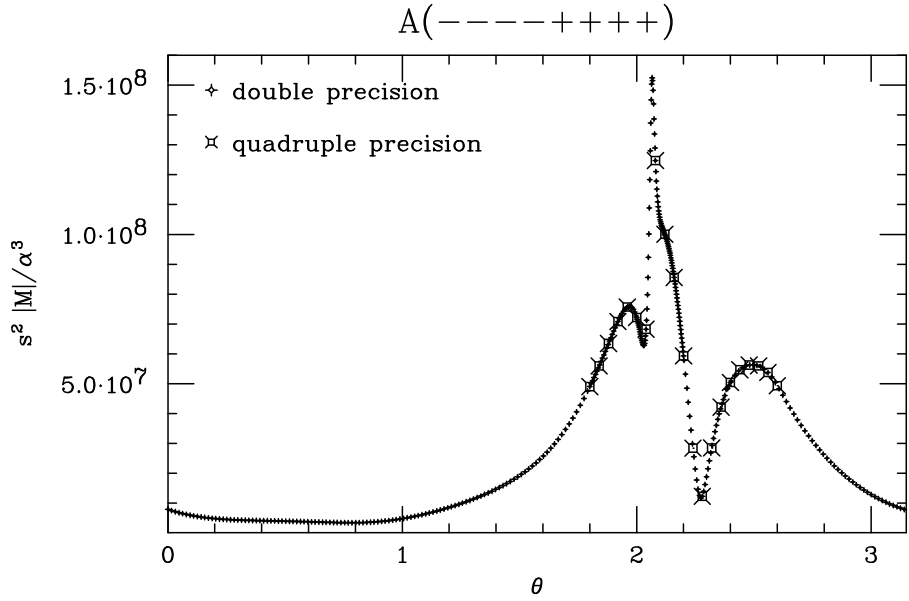}
\end{center}
\caption{
Results for the 8-photon amplitude with helicity $----++++$,
produced with {\samurai} in double-precision and with {\tt istop=4}.
21 points are also given with {\samurai} 
in quadruple-precision and with {\tt istop=2.}
}\label{fig:8photnoNNMHV}
\end{figure}

\noindent
The eight-photon amplitudes~\cite{Mahlon:1993fe,Gong:2008ww,Badger:2008rn}
are an example of the functionality of {\samurai} for 
many-particle scattering. \\
The numerator function is written along the same lines as in the previous two
sections. In this case, the number of diagrams is 5040. 
We evaluate the amplitudes for two helicity choices.\\
By using the same sampling set as in \cite{Gong:2008ww},  
we show in Fig.\ref{fig:8photnoMHV} how 
the numerical result produced with {\samurai} in the MHV case, $--++++++$, 
are tight to the analytic behavior \cite{Mahlon:1993fe}. 
The NNMHV case, $----++++$, shown in Fig.\ref{fig:8photnoNNMHV}, 
is a new result that confirms the structure of 
the amplitude discussed in \cite{Badger:2008rn},
where only boxes do contribute.

\subsection{Drell-Yan}

\begin{figure}[h]
\begin{center}
\includegraphics[scale=0.7]{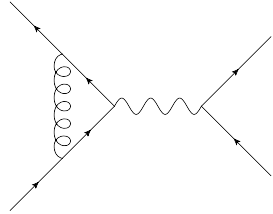}
\caption{Triangle diagram for Drell-Yan.}\label{fig:Triangle}
\end{center}
\end{figure}

The one-loop correction to $u \bar u \to e^+ e^-$ \cite{Altarelli:1978id,Altarelli:1979ub}
is an easy example of a numerator with $\epsilon$-dependent terms.
The numerator of the diagram in Fig.~\ref{fig:Triangle}
can be cast in the form
\bea
N(q, \mu^2) &=& C_F\,g_s^2\,e^2\,\bar{u}(p_{e^-})\,\gamma^\mu\,v(p_{e^+})\,
                \bar{v}(p_{\bar{u}})\,\big[\,2\,(2-d)\,\bar{q}^\mu\,\slh{\bar{q}}\,+\,
                [\,(d-2)\,\bar{q}^2 \nn
            & & +4\,(p_u \cdot \bar{q} - p_{\bar{u}} \cdot \bar{q} - p_u \cdot p_{\bar{u}})\,]\,
                \gamma^\mu\,\big]\,u(p_u) \nonumber
\eea
with denominators
\beq
\bar{q}^2 \,\,\,\,\, (\bar{q}+p_u)^2 \,\,\,\,\, (\bar{q}+p_u+p_{e^-}+p_{e^+})^2 \,. \nonumber
\eeq
The value $d=4$ in the expression above corresponds to the result 
in the Dimensional Reduction (DR) scheme, while 
the choice $d=4-2\epsilon$ yields an $\epsilon$-dependent
term, according to the Conventional Dimensional Regularization (CDR) scheme.
{\samurai} can be used to reduce both the $\epsilon^0$ 
and the coefficient of the
$\epsilon^1$ term individually, namely $N_0$ and $N_1$ of Eq.(\ref{eq:genericN}).
It is easy to see that the inclusion of the latter has the
well known effect of subtracting a contribution $C_F\,g_s^2$ times 
the tree-level amplitude from the finite part of the DR-result. 

\subsection{Leading-color Amplitude for $V+1$jet}
\begin{figure}[t]
\begin{center}
\includegraphics[scale=0.7]{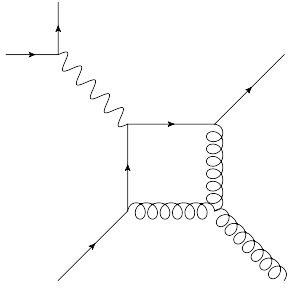}
\caption{Parent diagram for the amplitudes $A_{5;1}$.}\label{fig:A51pardiag}
\end{center}
\end{figure}
The leading color amplitude for the virtual NLO correction
to $V+1$jet production at the hadron collider is a good exercise
to show the reduction in a case where the contribution of all diagrams 
is cast in a single numerator function.

Once the color factors have been stripped, 
this amplitude can be calculated at the
Feynman diagram level taking the sum of the parent diagram in Fig.~\ref{fig:A51pardiag}
and its pinched diagrams, i.e. four triangles and two bubbles.
The presence of the $\gamma^5$ in the weak vertex imposes a choice on its
treatment in dimensional regularization. Adopting the Dimensional Reduction (DR)
scheme and assuming an anticommuting $\gamma^5$ one can get the right result
adding a well known finite-renormalization contribution, 
amounting to $(-N_c/2)$ times the tree-level amplitude.

With the proper routing of the loop momentum in the diagrams, it is possible
to collect all the diagrams over the four denominators of the parent box: 
the numerator of triangles is multiplied by the single missing denominator,
while the bubbles by two denominators. 
In this, we should process only one numerator function. 
This way of collecting the diagrams does not spoil the precision of the result. Using this construction, we found perfect agreement with the expression for $A_{5;1}$ given in the Eqs.(D.1-D.5) of ~\cite{Bern:1997sc}.

\subsection{Five- and Six-gluon amplitudes}

We choose two simple examples, namely the amplitudes contributing 
to the rational part of the {\it all-plus helicity}
5-gluon and 6-gluon scattering 
\cite{Bern:1993sx,Bern:1993qk,Brandhuber:2005jw,Badger:2008cm},
to show how a unitarity-based calculation can be implemented 
within {\samurai} (option {\tt imeth=tree}). 

\begin{figure}[h]
\begin{center}
\includegraphics[scale=0.7]{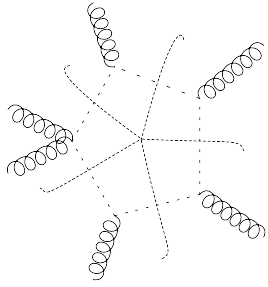}
\caption{Quintuple-cut of the 6-gluon amplitude.}\label{fig:6gcut5}
\end{center}
\end{figure}

The diagrams involved correspond to one-loop amplitudes 
with external gluons coupled 
to a massive-scalar loop, whose 
integrand can be built by means of the tree-level amplitudes 
given in \cite{Brandhuber:2005jw,Badger:2005zh}, namely
\bea
A_3^{\rm tree}(1_s; 2^+; 3_s) \!&\!=\!&\!
{\spba 2.1.{r_2} \over \spa 2.{r_2}} \ , \\ 
A_4^{\rm tree}(1_s; 2^+, 3^+; 4_s) \!&\!=\!&\!
 {\mu^2 \ \spb 2.3 \over \spa 2.3 (p_{12}^2 - \mu^2)} \ , \\ 
A_5^{\rm tree}(1_s; 2^+, 3^+, 4^+; 5_s) \!&\!=\!&\!
 {\mu^2 \ \spbb 2.1.{(2+3)}.4 \over 
       \spa 2.3 \spa 3.4 
       (p_{12}^2 \!-\!\mu^2)
       (p_{45}^2 \!-\! \mu^2)} \ , \quad
\eea
where $r_2$ is the reference vector of the gluon-2,
and $p_{ij} = k_i + k_j$. For instance,
the integrand of the quintuple-cut shown in Fig.\ref{fig:6gcut5} 
can be written as,
\bea
N(q,\mu^2) \!\!&=&\!\! 
               A_4(L_1; 1^+,2^+; -L_2) \times 
               A_3(L_2; 3^+;-L_3) \times 
               A_3(L_3; 4^+;-L_4) \qquad \nn && \times 
               A_3(L_4; 5^+;-L_5) \times 
               A_3(L_5; 6^+;-L_1) 
\eea
where
\bea
L_1 = q \ , \ 
L_2 = q + p_{12} \ , \ 
L_3 = q + p_{123} \ , \
L_4 = q + p_{1234} \ , \ 
L_5 = q - p_6 \ . 
\eea
In the case of the 5-gluon amplitudes,
we give the complete set of integrands, 
for quintuple-, quadruple-, triple- and double-cuts
({\tt istop=2}), although only boxes appear in the result.
In this case, we see explicitly that triangles and bubbles 
do not contribute.
For the same reason, in the 6-gluons case, we only give the integrands 
for the quintuple- and quadruple-cuts ({\tt istop=4}).

The results of these calculations, due to the external helicity choice, 
are purely rational in the $d=4$ limit and agrees with the results of 
\cite{Badger:2005zh}.

\subsection{Six-Quarks Scattering}
\label{sec:sixquarks}

When the number of diagrams contributing to the scattering amplitude is small, 
the input file that includes the numerators and the list of momenta to be processed by the reduction is fairly simple, and the calculations are feasible
with a minimal amount of automation~\cite{Binoth:2008kt,Actis:2009uq}.
However, even in simple cases, a careful automation reduces the probability of
introducing bugs or human mistakes in the code.

An automatized generation of the input files becomes a necessity as the complexity of the process increases.
As a final example (with the diagrammatic approach), 
we tackle a more involved calculation, 
namely the one-loop QCD corrections to the 6-quark scattering $q_1 \bar q_1 \to 
q_2 {\bar q}_2 q_3 {\bar q}_3$. The number of Feynman diagrams contributing to this process requires a 
fully automated approach. \\
The amplitude for  $q_1 \bar q_1 \to q_2 {\bar q}_2 q_3 {\bar q}_3$ involves 258 Feynman diagrams (8 hexagons, 24 pentagons, 42 boxes, 70 triangles, and 114 bubbles). Each diagram, or convenient combinations of them, should be processed by the 
reduction algorithm separately. 
The numerators and the lists of denominators required by the reduction have been generated and automatically written in a {\tt Fortran90} code, ready to be processed by {\samurai}. 

We use this example also as a first benchmark on the functionality of our framework. 
During the generation of the code, 
all Feynman diagrams contributing to the process are automatically written 
and organized in {\tt Fortran90} files fully compatible with the reduction library, 
ready to be run. \\
In order to check our algebraic manipulations, we compute both $N_0(q)$ and $N_1(q)$ of Eq.(\ref{eq:genericN}), namely also the part of the numerator proportional to $\epsilon$, although in an actual calculation this can be avoided by choosing the regularization scheme conveniently.

There are eight different helicity configurations that contribute to this process. 
Our numerical results have been compared with those obtained for the same process with {\tt golem-2.0} and {\tt golem95}~\cite{Binoth:2010pb} and we found perfect agreement.

On a Intel(R) Xeon(R) CPU X5482 3.20GHz machine,
the generation of the code for the full process takes less than 10 minutes,
and the result for each color summed helicity amplitude is produced in 55 ms per phase-space point. 
However, by avoiding the reduction of $N_1(q)$ with a proper scheme choice, 
the computing time goes down to 36 ms/ps-point.

\subsubsection{Numerator}
\label{sec:numerator}

\begin{figure}[tbh]
\begin{center}
\includegraphics[scale=0.7]{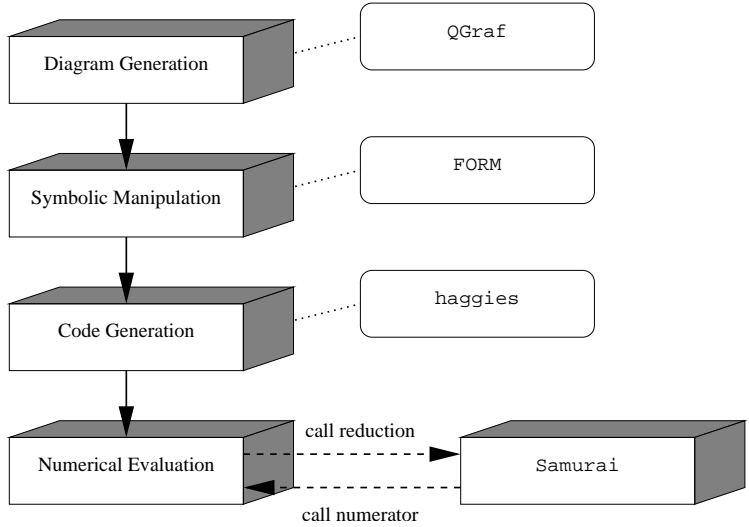}
\end{center}
\caption{Schematic overview of the generation of the numerator.
The boxes correspond to the steps required for the generation of
the numerator for a given process using Feynman diagrams.
The actual implementations we have used for each step are given in rounded
rectangles (see text). The dashed arrows indicate control flow, plain arrows
indicate data flow.}\label{fig:workflow}
\end{figure}

When working with Feynman diagrams, we prepare the numerator
function $\mathcal{N}(\bar{q})$ by processing the output of a
diagram generator symbolically with a computer algebra program;
the actual computer program is written by an optimizing code generator
(see also \fig{fig:workflow}).
This modular approach is very generic and, to a large extend, can be based
on existing tools; in particular we have an automated setup using
\texttt{QGraf}~\cite{Nogueira:1991ex}, \texttt{Form}~\cite{Vermaseren:2000nd}
and \texttt{haggies}~\cite{Reiter:2009ts}.
Furthermore, the matrix element generator
\texttt{golem-2{.}0}~\cite{Binoth:2010pb} has been extended to provide
an interface which simplifies the use of the components mentioned above.
We want to stress that the described setup is very modular and that
any component in the workflow can be exchanged by alternative solutions.

As discussed in Section~\ref{sec:Reduction}, the most general numerator of one-loop amplitudes,
$\mathcal{N}(\bar{q}, \epsilon)$, can be written as,
\begin{equation}
\mathcal{N}(\bar{q}, \epsilon) =
            N_0(\bar{q})
+ \epsilon  N_1(\bar{q})
+\epsilon^2 N_2(\bar{q}).
\end{equation}
The functions $N_0$, $N_1$ and $N_2$ are functions of
$q^\nu$ and $\mu^2$, therefore in our discussion,
except when it is necessarily required a distinction between them, 
we will simply talk about $N$, giving as understood that the
same logic would apply to each of the three contributions~$N_i$.

We work with the helicity projections of the amplitude
which are decomposed into subamplitudes formed by the sum of all diagrams
sharing the same set of denominators. The color information is hidden
from the reduction by defining the numerators of the
subamplitudes, $\mathcal{N}^{(i)}(\bar{q},\epsilon)$,
as the contraction of the numerators of the one-loop diagrams with the
tree-level amplitude.
If we call $N_{\{i_1i_2\ldots i_n\}}$ the numerator stemming from the
sum of all diagrams
which have (exactly) the denominator $\db{i_1}\db{i_2}\cdots \db{i_n}$,
the corresponding
subamplitude would~be
\begin{equation}
\mathcal{N}^{(i)}(\bar{q},\epsilon)=
\mathcal{A}_{\text{born}}^\dagger\cdot
N_{\{i_1i_2\ldots i_n\}}.
\end{equation}
In our implementation this product is done numerically and does
therefore not add to the complexity of the expressions.
In cases where the tree-level matrix element vanishes, one can always
find an appropriate set of color
projectors%
~$\mathcal{P}_I^\dagger\mathcal{P}_I$ into one-dimensional subspaces
such that
\begin{equation}
\mathcal{A}_n^\dagger\cdot\mathcal{A}_n=
\sum_I (\mathcal{P}_I\mathcal{A}_n)^\dagger\cdot
(\mathcal{P}_I\mathcal{A}_n)\text{.}
\end{equation}
where $\mathcal{P}_I$ correspond to Wigner-Eckhard symbols.
In the cases with no external color, the only projection is~$\mathcal{P}_0=1$.
The objects $\mathcal{P}_I\cdot N_{\{i_1i_2\ldots i_n\}}$
hence are the objects that undergo the reduction.


Optionally, one can also group larger
sets of diagrams into subamplitudes by also considering diagrams which contain
a subset of the maximal set of denominators.
The numerator of the corresponding
subamplitude in the latter sense would~be
\begin{multline}
\mathcal{N}^{(i)}=\left[N_{\{i_1i_2\ldots i_n\}}
+\db{i_n}N_{\{i_1i_2\ldots i_{n-1}\}}+\db{i_{n-1}}N_{\{i_1\ldots i_{n-2}i_n\}}
+\ldots\right.\\\left.
+\db{i_1}\db{i_2}\cdots \db{i_{n-2}}\db{n}N_{\{i_{n-1}\}}
+\db{i_1}\db{i_2}\cdots \db{i_{n-1}}N_{\{i_n\}}%
\right]\cdot\mathcal{A}_{\text{born}}^\dagger.
\end{multline}

\subsubsection{Algebraic Simplification of the Lorentz Structure}
In order to unravel the dependence of $\mathcal{N}(\bar{q},\epsilon)$
on $q$, $\mu^2$ and~$\epsilon$ we use dimension splitting based on
the 't~Hooft-Veltman scheme.
We define the subspaces of the regulated Minkowski space such that
\begin{equation}
\bar{g}^{\mu\nu}=g^{\mu\nu}+\tilde{g}^{\mu\nu},\quad
\bar{g}^\mu_\mu=d,\quad g^\mu_\mu=4,\quad \tilde{g}^\mu_\mu=-2\epsilon,
\quad g^{\mu\rho}\tilde{g}_{\rho\nu}=0
\end{equation}
and with the corresponding projections of the Dirac matrices
$\gamma^\mu=g^\mu_\nu\bar{\gamma}^\nu$ and
$\tilde{\gamma}^\mu=\tilde{g}^\mu_\nu\bar{\gamma}^\nu$ the
Dirac algebra is uniquely defined by
\begin{equation}
\{\bar{\gamma}^\mu,\bar{\gamma}^\nu\}=2\bar{g}^{\mu\nu},\quad
\{\gamma^\mu,\gamma_5\}=0,\quad
[\tilde{\gamma}^\mu,\gamma_5]=0,\quad
\end{equation}

Working with this scheme one can show \cite{Reiter:2009kb}
that after separating the
four from the $(d-4)$ dimensional projection of each Dirac matrix
one can factorize a mixed spinor line into
\begin{multline}\label{eq:splitspinorline}
\langle p_\lambda\vert
\gamma^{\mu_1}\cdots\gamma^{\mu_k}%
\tilde{\gamma}^{\mu_{k+1}}\cdots\tilde{\gamma}^{\mu_l}%
\vert p^\prime_{\lambda^\prime}\rangle=
\langle p_\lambda\vert
\gamma^{\mu_1}\cdots\gamma^{\mu_k}%
\vert p^\prime_{\lambda^\prime}\rangle\cdot
\mathrm{tr}\{\tilde{\gamma}^{\mu_{k+1}}\cdots\tilde{\gamma}^{\mu_l}\}/%
\mathrm{tr}\{1\} \ .
\end{multline}
In this notation the definition of the helicities is such that
$\vert p_\pm\rangle=\frac12(1\pm\gamma_5)u(p)$ and
$\langle p_\pm\vert=\bar{u}(p)\frac12(1\pm\gamma_5)$,
where $p$ and $p^\prime$ are lightlike vectors.
The extension to massive vectors is straightforward by
projecting each massive vector onto a sum of two lightlike
vectors.
The trace in \eqn{eq:splitspinorline} evaluates to a product of
metric tensors $\tilde g^{\mu_i\mu_j}$
using the usual rules for spinor traces.
Since in the 't~Hooft-Veltman scheme
at one-loop the only $d$-dimensional vector is the integration momentum these
metric tensors lead to factors of $\mu^2$ and~$\epsilon$.
The Lorentz indices inside the remaining, four-dimensional spinor lines
are eliminated using Chisholm identities, of which we apply also a variant
specific to spinor chains,
\begin{equation}
\langle p_\lambda\vert\Gamma\gamma^\mu\Gamma^\prime%
\vert p^\prime_{\lambda^\prime}\rangle\cdot\gamma_\mu=
2\left(
\Gamma^\prime\vert p^\prime_{\lambda^\prime}\rangle%
\langle p_{\lambda}\vert\Gamma
-\lambda\lambda^\prime
\overleftarrow{\Gamma}\vert p_{\lambda}\rangle%
\langle p^\prime_{\lambda^\prime}\vert\overleftarrow{\Gamma}^\prime
\right)
\end{equation}
where $\Gamma$ and $\Gamma^\prime$ are strings of four-dimensional
Dirac matrices and $\overleftarrow{\Gamma}$ denotes the string in
reversed order. \\
After these steps, the numerator is suitable for efficient
numerical evaluation since it is expressed entirely in terms of
constants, dot products and spinor products of the form
$\langle p_\lambda\vert p^\prime_\lambda\rangle$ and
$\langle p_\lambda\vert\slh{q}\vert p^\prime_{-\lambda}\rangle$.

\subsubsection{Result of ${\cal A}(1^-,2^+,3^-,4^+,5^-,6^+)$}

\bigskip

The LO contribution and the NLO virtual corrections to the
squared amplitude (ultra-violet renormalised) 
are defined as,
\begin{align}
a_{\rm LO} &= \mathcal{A}^\dagger_{\text{LO}}\mathcal{A}_{\text{LO}}\\
\mathcal{A}^\dagger_{\text{virt}}\mathcal{A}_{\text{LO}}+h.c.&=
a_{\rm LO}\cdot
\frac{\alpha_s}{2\pi}\frac{(4\pi)^\epsilon}{\Gamma(1-\epsilon)}
\left(
\frac{a_{-2}}{\epsilon^2}
+\frac{a_{-1}}{\epsilon^1}
+a_{0}
\right)
\end{align}

\noindent
The result of \texttt{golem95} for the helicity configuration 
$(q_1^-,{\bar q}_1^+,q_2^-,{\bar q}_2^+,q_3^-,{\bar q}_3^+)$,
at the ps-point given in Eq.(\ref{eq:6photonpsp}), is
\begin{align}
a_{\rm LO}&=0.9686295685264447\times10^{-6} \ ,\\
a_{-2}&=-8.000000000048633 \ ,\\
a_{-1}&=46.40675046335535 \ ,\\
a_{0}&=-233.8908276457752 \ ;
\end{align}
and the one computed by {\samurai} is
\begin{align}
a_{\rm LO}&=0.9686295685264458\times10^{-6} \ ,\\
a_{-2}&=-7.999999999999935 \ ,\\
a_{-1}&=46.40675045992446 \ ,\\
a_{0}&=-233.8908276128404 \ ,
\end{align}
showing a nice agreement
(the color-average factor, $1/9$, and the helicity-average factor, $1/4$, are
already included). \\
The double- and single-pole of the virtual
contribution are consistent with the infrared poles amounting to \cite{Catani:1996vz},
\begin{align}
a_{-2}&=8.000000000000000 \ ,\\
a_{-1}&=-46.40675046319159 \ .
\end{align}

\subsubsection{Precision of Integrated Results}
\label{sssec:6qint}
\begin{figure}[ht]
\begin{center}
\includegraphics[width=0.8\textwidth]{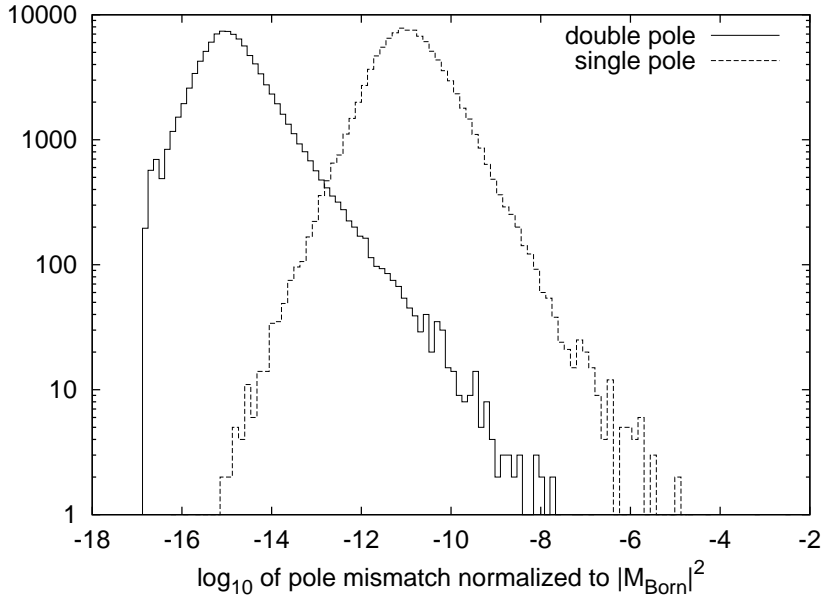}
\end{center}
\caption{
Estimate for the precision obtained from the difference between the single
(resp. double) poles of the virtual amplitude and those of the
integrated dipoles for
$q_1\bar{q}_1\rightarrow q_2\bar{q}_2q_2\bar{q}_2$.
The results have been obtained by integrating $10^5$ phase-space points at
$\sqrt{s}=14\,\text{TeV}$,
where we have used cuts on $p_T>30\,\text{GeV}$ and the rapidity $\eta<2.5$
as well as a separation cut of $\Delta R>0.8$ between the final state
particles. We used the CTEQ6m~\cite{Kretzer:2003it}
PDF set with two-loop running for $\alpha_s$ with a renormalisation scale
of $\mu=\sqrt{\sum_i p_T(i)^2}$.
}\label{fig:6qintpoles}
\end{figure}

We have used the matrix element of the
$q_1\bar{q}_1\rightarrow q_2\bar{q}_2q_3\bar{q}_3$ amplitude
for recalculating the
$q_1\bar{q}_1\rightarrow q_2\bar{q}_2q_2\bar{q}_2$ amplitude~\cite{Binoth:2010pb} by anti-symmetrizing over the final state.
We have integrated the virtual matrix element with MadEvent~\cite{Maltoni:2002qb,Alwall:2007st}
and compared the poles of the virtual amplitude to those of the integrated
dipoles using MadDipole~\cite{Frederix:2008hu,Frederix:2010cj}. Figure~\ref{fig:6qintpoles}
shows the remainder of the pole contributions
which should sum up to zero.
The results represent a realistic Monte Carlo integration and
indicate that the precision is well under control.

\section{Conclusions}
In this work we have presented
{\samurai}, a tool for the automated numerical evaluation 
of one-loop corrections to any scattering amplitudes within 
the dimensional-regularization scheme. 
Its implementation is based on the decomposition of the integrand according to 
the OPP-approach, extended to the framework of the generalized $d$-dimensional 
unitarity-cuts technique, and on the use of the 
Discrete Fourier Transform as polynomial interpolation technique. 
We have shown how {\samurai} can process 
integrands written either as numerator 
of Feynman integrals, like in diagrammatic methods, 
or as product of tree-level amplitudes, according to unitarity-based methods. 
In both cases, the advantage of working within a $d$-dimensional 
unitarity framework 
is that the result of {\samurai} 
is complete and does not require any additional
information for the reconstruction of the rational terms.

We discussed its application to a series of examples such as 
the 4-, 6-, and 8-photon scattering amplitudes in QED, 
the QCD virtual corrections to Drell-Yan, 
the leading color amplitude for $V+1$jet production, 
the six-quark amplitudes, 
and contributions from massive-scalar loop to 
the all-plus helicity 5- and the 6-gluon amplitudes.
For the six-quark scattering
$q_1 \bar q_1 \to q_2 {\bar q}_2 q_3 {\bar q}_3$,
we also considered a fully automated reduction, from the integrand 
generation to the final result.

Given the versatility of the code,
{\samurai} may constitute a useful module 
for the systematic evaluation of the virtual corrections,
oriented towards the automation of next-to-leading order calculations 
relevant for the LHC phenomenology.

\bigskip

The reduction library {\tt libsamurai} and the examples are 
publicly available at 
the webpage:
\begin{center}
{\tt http://cern.ch/samurai}
\end{center}

\section*{Acknowledgments}

We are indebted to Simon Badger for feedback and comparisons on the 5- and 
6-gluon amplitudes, and on the MHV 8-photon amplitudes.
We also thank Jean-Philippe Guillet for 
the numerical comparisons of the 6-photon amplitudes.
We like to thank Nicolas Greiner for providing the MadEvent
code used for the example in Section~\ref{sssec:6qint}.
P.M. and F.T. are pleased to thank Zoltan Kunszt, 
Zoltan Trocsanyi and Bryan Lynn for clarifying discussions.
G.O. and T.R. wish to acknowledge the kind hospitality of the Theory Department at CERN, 
at several stages while this project has been performed.
The work of G.O. was supported by the NSF Grant PHY-0855489 and PSC-CUNY Award~60041-39~40; T.R. has been supported by the Foundation FOM, project FORM 07PR2556.

\bibliographystyle{utphys} 
\bibliography{MasterMORT.bib}
%
\end{document}